\documentclass[preprint,1p]{elsarticle}
\usepackage{graphicx}
\usepackage{amsmath}
\usepackage{natbib}
\usepackage[usenames]{color}
\usepackage{soul}

\usepackage[normalem]{ulem}

\usepackage[utf8]{inputenc}

\usepackage{amsfonts}
\usepackage{dcolumn}        
\usepackage{amssymb}
\usepackage{bm,amssymb}

\usepackage{bm,fancybox}                	     

\begin{document}

\newcommand{\bea}{\begin{eqnarray}}
\newcommand{\eea}{  \end{eqnarray}}
\newcommand{\bit}{\begin{itemize}}
\newcommand{\eit}{  \end{itemize}}

\newcommand{\be}{\begin{equation}}
\newcommand{\ee}{\end{equation}}
\newcommand{\ra}{\rangle}
\newcommand{\la}{\langle}
\newcommand{\U}{\widetilde{U}}

\def\bra#1{{\langle#1|}}
\def\ket#1{{|#1\rangle}}
\def\bracket#1#2{{\langle#1|#2\rangle}}
\def\inner#1#2{{\langle#1|#2\rangle}}
\def\expect#1{{\langle#1\rangle}}
\def\e{{\rm e}}
\def\proj{{\hat{\cal P}}}
\def\tr{{\rm Tr}}
\def\H{{\hat H}}
\def\Hdag{{\hat H}^\dagger}
\def\Lop{{\cal L}}
\def\Ehat{{\hat E}}
\def\Edag{{\hat E}^\dagger}
\def\Shat{\hat{S}}
\def\Sdag{{\hat S}^\dagger}
\def\Ahat{{\hat A}}
\def\Adag{{\hat A}^\dagger}
\def\U{{\hat U}}
\def\Udag{{\hat U}^\dagger}
\def\Zhat{{\hat Z}}
\def\Phat{{\hat P}}
\def\Op{{\hat O}}
\def\id{{\hat I}}
\def\x{{\hat x}}
\def\P{{\hat P}}
\def\Px{\proj_x}
\def\Pr{\proj_{R}}
\def\Pl{\proj_{L}}


\title{Average localization of resonances on the quantum repeller}
\author[add1]{J. Montes}
 \ead{jmontes.3@alumni.unav.es}

\author[add2]{Gabriel G. Carlo}
\ead{carlo@tandar.cnea.gov.ar}

\author[add1]{F. Borondo}
 \ead{f.borondo@uam.es}

 \address[add1]{Departamento de Qu\'imica, 
 Universidad Aut\'onoma de Madrid,
 Cantoblanco, 28049--Madrid, Spain}

 \address[add2]{Comisi\'on Nacional de Energ\'ia At\'omica, CONICET, Departamento de F\'isica, 
 Av.~del Libertador 8250, 1429 Buenos Aires, Argentina}

\date{\today}

\begin{abstract}
There has been a very recent surge in the interest on the localization properties of resonances 
associated to partially open (scattering) systems, which are of great relevance when 
studying resonant cavities such as those used in microlasers. 
Very recently, it has been found that 
no localization is present in a scaled form of these states. 
Moreover, a new kind of scarring on structures different from periodic orbits 
is described for non scaled resonances. 
In this paper, we analyze the localization of a distribution function corresponding to the quantum 
LR representation --based on the non unitary evolution operator decomposition into left and right resonances-- 
for the partially open quantum tribaker map, a paradigmatic system. 
We find localization on the shortest periodic orbits. 
Also, scaled states present enhancements that could 
not be associated to periodic orbits and that become more evident 
when looking at the LR representation. 
These findings open the door for new perspectives on recent theoretical developments. 
\end{abstract}
\maketitle

\section{Introduction}
 \label{sec:intro}

 Quantum scattering has been an active research field since long ago, 
 being optical cavities one of the main focus of attention. 
There is a long history of both theoretical and experimental studies \cite{Cao,Novaes},
but also recent experiments \cite{RecentExp} 
in which a lot of questions have been settled, but others still remain open.
 This situation corresponds classically to placing an \textit{opening} in the system,
 letting the trajectories that arrive at a region of phase space escape. 
 This escape can be complete, and then we speak about a complete opening, 
 or partial, where we define a convenient reflectivity function in order to take into account 
 a partial confinement of these trajectories inside the cavity. 
 This mechanism gives rise to a classical (conditionally) invariant 
 distribution that corresponds to a (partial) fractal set having (multi) fractal dimension(s). 
 At the quantum level and for the complete opening case, the fractal Weyl law is a well
 established result, which shows that the number of long-lived states (or resonances) scales 
 with the  Planck constant as $\hbar^{-d/2}$, where $d+1$ is the fractal dimension of the 
 classical \textit{repeller} \cite{conjecture,Nonnenmacher,Hamiltonians,qmaps}. 
 If the opening is partial \cite{Blumel}, we have multifractal measures \cite{Ott} that characterize 
 a conditionally invariant distribution that extends over all the phase space and the usual fractal 
 Weyl law has to be redefined. 
 This has been done for the very important case of maps 
 (these systems embody all the features of the chaotic cavities, but are much more amenable to 
 theoretical studies) \cite{Altmann}. 
 The morphology of the corresponding eigenfunctions is still not completely understood, 
 being their localization properties \cite{Dyatlov} a source of active developments nowadays.
 
 Very recently there has been a renovated interest in elucidating some open questions about the structure of resonances associated to (partially) open systems. 
 In \cite{Ketz1} the average phase space distribution of resonances in chaotic systems 
 with escape was conjectured to be described by a classical measure. 
 This measure is conditionally invariant and uniformly distributed on sets with the same temporal 
 distance to the quantum resolved chaotic saddle. 
 A family of conditionally invariant classical measures was found to describe the multifractal 
 phase space distribution, product structure along stable and unstable directions, 
 and the dependence on the decay rate of resonances \cite{Ketz2} 
 (better behaved for long and short lived ones). 
 Moreover, in \cite{Ketz3} the intensity statistics was found to universally follow 
 an exponential distribution in the partially open chaotic standard map, baker map, 
 and a random matrix model. 
 In \cite{Ketz4} a local randomization on phase space for the baker map with escape was 
 introduced to obtain a semiclassical description of resonances, though some quantitative differences with 
 respect to the exact ones were found. 
 Finally, in \cite{Ketz5} the resonances were conjectured to be a product of an 
 essentially classical conditionally invariant measure and universal fluctuations. 
 Notably, scarring is present in almost all of them, although along segments of rays but not along periodic orbits (POs).
 
 On the other hand, there has been a complementary line of research in order 
 to study the properties of resonances based on the localization phenomena on POs, the so called scarring \cite{art0}. 
 This led to the development of the semiclassical theory of short POs  for open quantum maps \cite{art1}. 
 This is a constructive approach based on short POs contained in the 
classical repeller, which motivates a basis of scar functions able to expand the quantum repeller
\cite{art2,art3}, which is suitable to express the quantum non-unitary operators
\cite{art4,art5}. We notice that the term short refers to the fact that their periods are related to the number 
of periodic points needed to cover the quantized torus in the case of maps, and as a consequence 
their length is much shorter than that of the orbits needed in other semiclassical theories, such as the trace 
formula for example.
We have recently extended the short POs theory to partially open quantum maps \cite{art6,art7}. 

Can these two visions on the same problem be put together? 
Is scarring in the long lived resonances of partially open systems 
essentially different from that occurring along POs in the closed scenario? 
How are the limits corresponding to completely open and closed systems reached?
Which is the effect of the scaling of these resonances by their average 
in terms of localization?  
In this work we try to answer these questions by studying the 
localization properties of the longest lived resonances of the (partially)
open tribaker map. 
We use the Husimi functions associated to the right eigenvectors 
scaled using their average, and the \textit{LR} representation of the left and 
right eigenvectors scaled using the corresponding \textit{quantum repeller}. 
In this way, we are able to clearly show the localization present when using 
the latter mathematical objects, which reveal themselves as a very adequate 
tool to understand the statistical properties of the resonances. 
Moreover, thanks to them scarring on short POs is easily identified.
 
The organization of this paper is as follows: 
in Sec.~\ref{sec:SystemandMeasures} we describe the (partially) open tribaker map 
and the localization measures used. 
In Sec.~\ref{sec:Localization} we apply these definitions to unveil the localization properties 
of resonances and how short POs could be apparently hidden in the set of long lived resonances. 
The conclusions are outlined in Sec.~\ref{sec:Conclusions}.

\section{System and measures}
 \label{sec:SystemandMeasures}

Maps are a powerful and convenient tool to study classical and quantum chaos 
\cite{Ozorio 1994,Hannay 1980,Espositi 2005}. 
In particular, open maps on the 2-torus have an associated fractal repeller,
that is an invariant measure which lies at the intersection of the forward and backwards 
trapped sets (nonescaping trajectories in the past or future). 
When the opening is partial, some amount of them is reflected following a reflectivity 
function that we will take as a constant $R \in (0:1)$ for simplicity. 
In this case, the corresponding measure extends over all the phase space, 
but still shows multifractality. 

The quantum version of maps on the torus requires taking 
$\bracket{q+1}{\psi}\:=\:e^{i 2 \pi \chi_q}\bracket{q}{\psi}$, and
$\bracket{p+1}{\psi}\:=\:e^{i 2 \pi \chi_p}\bracket{p}{\psi}$, with $\chi_q$, $\chi_p \in [0,1)$. 
The dimension of the Hilbert space is $N=(2 \pi \hbar)^{-1}$, the
semiclassical limit is reached when $N \rightarrow \infty$, and the evolution operator is a
$N\times N$ matrix. Position and momentum eigenstates are given by
$\ket{q_j}\:=\:\ket{(j+\chi_q)/N}$ and $\ket{p_j}\:=\:\ket{(j+\chi_p)/N}$ with
$j\in\{0,\ldots, N-1\}$, transforming as 
$\bracket{p_k}{q_j}\:=\: \frac{1}{\sqrt{N}} e^{-2i\pi(j+\chi_q)(k+\chi_p)/N} \: \equiv \:
(G^{\chi_q, \chi_p}_N)$. 
A partially open quantum map is non-unitary and has 
$N$ right eigenvectors $|\Psi^R_j\ra$ and $N$ left ones $\la \Psi_j^L|$ 
for each resonance (eigenvalue) $z_j$, 
with $\la \Psi_j^L|\Psi^R_k\ra=\delta_{jk}$ and 
$\la \Psi_j^R|\Psi^R_j\ra=\la \Psi_j^L|\Psi^L_j\ra$ being the norm.

The classical tribaker map is defined as
\begin{equation}
\mathcal B(q,p)=\left\{
  \begin{array}{ll}
  (3q,p/3) & \mbox{if } 0\leq q<1/3 \\
  (3q-1,(p+1)/3) & \mbox{if } 1/3\leq q<2/3\\
  (3q-2,(p+2)/3) & \mbox{if } 2/3\leq q<1\\
  \end{array}\right.
\label{classicaltribaker}
\end{equation}
If we place an opening in the region $1/3< q<2/3$ all trajectories arriving 
at it will suffer an intensity modification given by the reflectivity function. 
The quantum closed counterpart with antiperiodic boundary conditions 
$\chi_q=\chi_p=1/2$ (preserving time reversal and parity) in position representation is \cite{Saraceno1,Saraceno2}
\begin{equation}\label{quantumbaker}
 U^{\mathcal{B}}=G_{N}^{-1} \left(\begin{array}{ccc}
  G_{N/3} & 0 & 0\\
  0 & G_{N/3} & 0\\
  0 & 0 & G_{N/3}\\
  \end{array} \right).
\end{equation}
The corresponding partially open map is given by  
\begin{equation}\label{partialprojector}
 P=\left(\begin{array}{ccc}
  1_{N/3} & 0 & 0\\
  0 & \sqrt{R} \; 1_{N/3} & 0\\
  0 & 0 & 1_{N/3}\\
  \end{array} \right),
\end{equation}
 applied to Eq.~(\ref{quantumbaker}), in such a way that the original symmetries
 are preserved.

Motivated by the classical repeller, we can define the corresponding 
quantum object by means of an operator $\hat{h}_j$ constructed in terms of 
the right $\vert \Psi^R_j\rangle$ and left $\langle \Psi^L_j\vert$ eigenstates \cite{art2,art3}
\begin{equation}\label{eq.hdef}
 \hat{h}_j=\frac{\vert \Psi^R_j\rangle\langle \Psi^L_j\vert}{\langle \Psi^L_j\vert \Psi^R_j\rangle}.
\end{equation}
When the expectation values of these projectors 
are evaluated on coherent states $|q,p\rangle$, 
$\vert\langle q,p\vert \hat{h}_j\vert q,p\rangle\vert \propto 
\sqrt{{\cal H}^R_j (q,p) {\cal H}^L_j (q,p)}$, 
being ${\cal H}^{R,L}_j$ the Husimi functions of the $R, L$ resonances, 
which for closed (unitary) systems are simply the Husimi functions of the eigenstates. 
We define the quantum repeller as the average of the first $j$ of them, 
ordered by decreasing modulus of their eigenvalues  
($\vert z_l\vert\geqslant\vert z_m\vert$ with $l \leq m$): 
\begin{equation}
\hat{Q}_j\equiv \frac{1}{j} \sum_{j^\prime=1}^j\hat{h}_{j^\prime}, 
\label{Eq:repeller}
\end{equation}
and denote their phase space representation by means of coherent states as 
\begin{eqnarray}
 h_j(q,p)&=&\vert\langle q,p\vert \hat{h}_j\vert q,p\rangle\vert\\
 Q_j(q,p)&=& \vert\langle q,p\vert \hat{Q}_j\vert q,p\rangle\vert,
\end{eqnarray}
where $h_j(q,p)$ will be simply called the {\em LR representation} of the resonances. 
We have explicitly taken the modulus since these are complex functions in general. 
Also, for the sake of an easy comparison with the 
scaled Husimis that will be defined next, 
we give an equal weight to each resonance in the repeller sum,
although weighting them 
with the corresponding eigenvalues should provide a more faithful 
representation of the longest lived sector of the quantum map. 
In fact, in this work we are interested on this projector property of the quantum repeller, 
other possible choices for scaling like considering the moduli of each term in the 
sum of Eq. \ref{Eq:repeller} will be studied in the future.
It is worth noticing that, in this work, we focus on the long lived set of 
resonances and this sum runs on just a subset of the whole set. 
It is useful to define \cite{Ketz3} the scaled Husimi as
\[
 \tilde{{\cal H}}^{R}_i={\cal H}^{R}_i/{<\!{\cal H}^{R}\!>}_j
\]
and the scaled LR representation functions as 
\[
\tilde{h}_i=h_i/Q_j
\]
respectively, where the average $< \cdots >$ is taken over this $j$ subset of eigenstates. 

We explore two measures of localization for both scaled functions. 
The first one closely follows the conjecture of universal intensity statistics formulated in \cite{Ketz3}, i.e. that these functions obey an exponential distribution 
$P(w) dw = e^{- w} dw$, where $w=\tilde{{\cal H}}^{R}_i(q,p)$ or $w=\tilde{h}_i(q,p)$. 
We define the corresponding localization measure 
\[
\Sigma=\sum_{i}^{M_i} \|P(w_i) - e^{- w_i}\|,
\]
where $M_i$ is the number of intervals into which we divide the histogram $P(w_i)$ and $w_i$ is the value 
at which we evaluate the exponential function. This gives the deviation from the exponential behavior as 
the distance between both curves, for each case. 

The second measure of localization is intrinsically phase space motivated.
In fact, the norm ratio $\mu$ is taken to be \cite{art3}
\begin{equation}\label{eq.normratio}
 \mu(\tilde{h}_i)=\left( \frac{{\|\tilde{h}_i\|}_1/{\|\tilde{h}_i\|}_2}{{\|\rho_c\|}_1/{\|\rho_c\|}_2} \right)^2.
\end{equation}
We use $\tilde{h}_i(q,p)$ in its definition but is also valid for $\tilde{{\cal H}}^{R}_i(q,p)$. 
A coherent state at an arbitrary position $(q,p)$ in phase space is the 
normalization factor $\rho_c= \vert q,p \rangle \langle q,p \vert$, with the phase space norm given by
\begin{equation}\label{eq.phasenorm}
 {\|\tilde{h}_i\|}_\gamma=\left( \int_{{\cal{T}}^2} {\tilde{h}_i(q,p)^\gamma dq dp } \right)^{1/\gamma}.
\end{equation}
The norm ratio is also independent of the $h$ normalization, with a 
minimum value of $1$ for a maximally localized distribution (a coherent state) 
and a maximum value of $N/2$ for the uniform distribution.

\section{Localization in phase space}
 \label{sec:Localization}

\begin{figure}
\includegraphics[width=0.90\columnwidth]{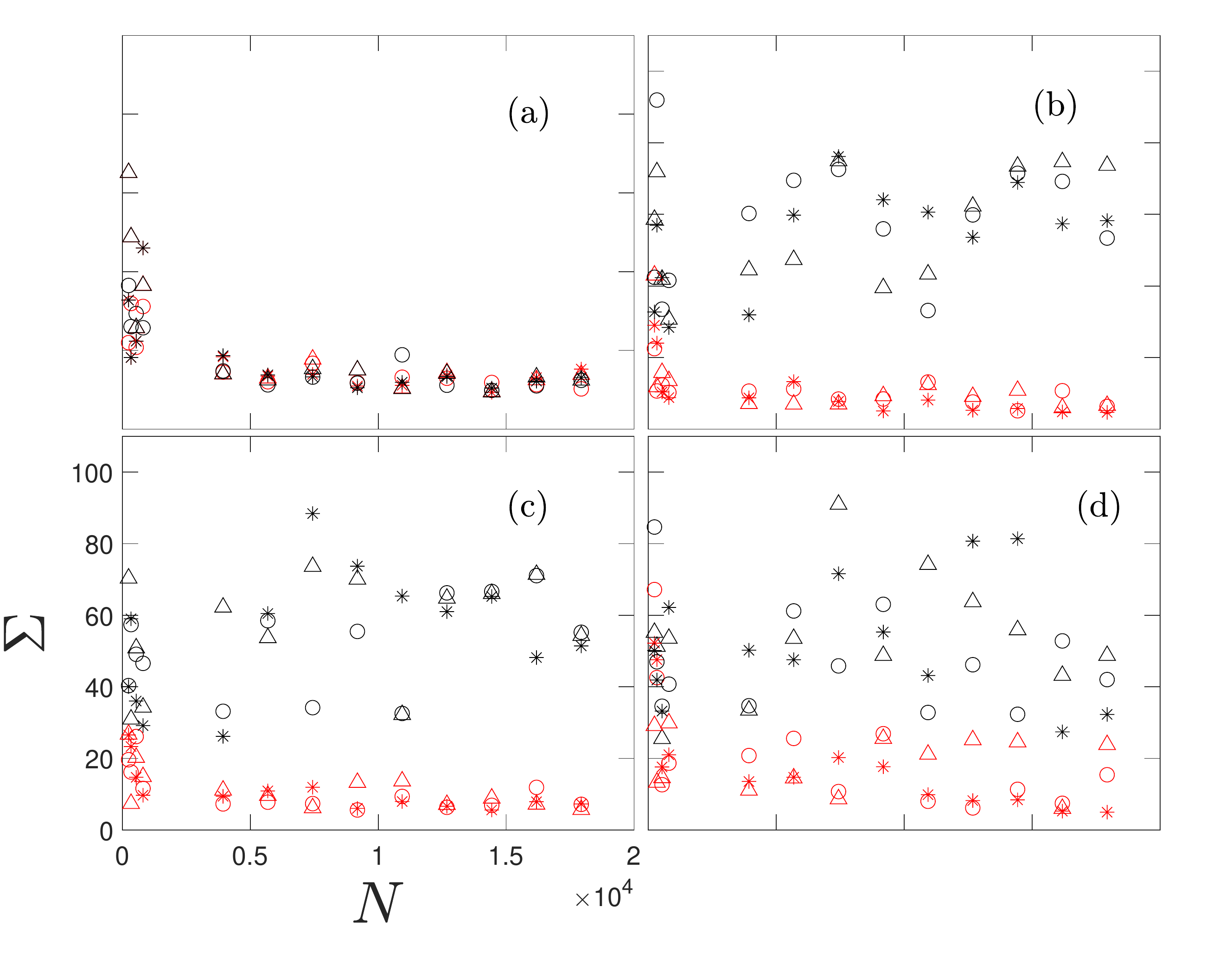}
\caption{(Color online) $\Sigma$ as a function of $N$ for the closed (a), 
partially open with $R=0.1$ (b), $R=0.05$ (c), and completely open (d) tribaker maps. 
The symbols $*$, $\triangle$ and $\circ$ stand for the 6th, 10th and 16th eigenstates (ordered as indicated in the main text), respectively. 
The gray (red) points correspond to the measure calculated for $\tilde{{\cal H}}^{R}_i$, 
while the black points for $\tilde{h}_i$.
}
\label{fig1}
\end{figure}

In order to investigate the average localization on the quantum repeller 
defined in Sec.~\ref{sec:SystemandMeasures}, 
we have computed the two previously defined measures,
$\Sigma$ and $\mu$, for the closed ($R=1$), 
partially open (with $R=0.1$ and $0.05$), and the completely open ($R=0$) tribaker maps. 
In the calculations, we have used the 32 longest lived resonances to evaluate 
$\tilde{{\cal H}}^{R}_i$ and $\tilde{h}_i$, which in the closed case are arbitrarily selected. 
Results for $\Sigma$ (calculated for $1000$ points distributions and 3 values of $i$) 
are shown in Fig.~\ref{fig1}.
%
Looking at it we verify that the exponential shape of the probability distributions is not 
a uniformly universal feature, but depends on the kind of representation used. 
For the closed system, there is essentially no difference 
between the Husimis of the right eigenstates and the distributions taken on the repeller, 
as expected from the fact that there are no differences between the right and left eigenstates. 
More remarkably, we see meaningful differences with respect to the 
localization for the closed system in the case of the repeller of the partially open maps, 
and also for the complete opening. 
The low to moderate $R$ values that have been considered for the partially open cases produce 
appreciable changes in the behavior of this measure with respect to the closed case. 
It needs to be underlined that for the Husimi distributions of the right eigenstates 
there is no notable change from the closed scenario at any of the finite reflectivities shown. 
This agrees with the universality conjecture of \cite{Ketz3} 
and shows that our calculations are in that regime. In fact, the number of 
resonances that we have considered in performing the scaling shows to be adequate.

\begin{figure}
\includegraphics[width=0.90\columnwidth]{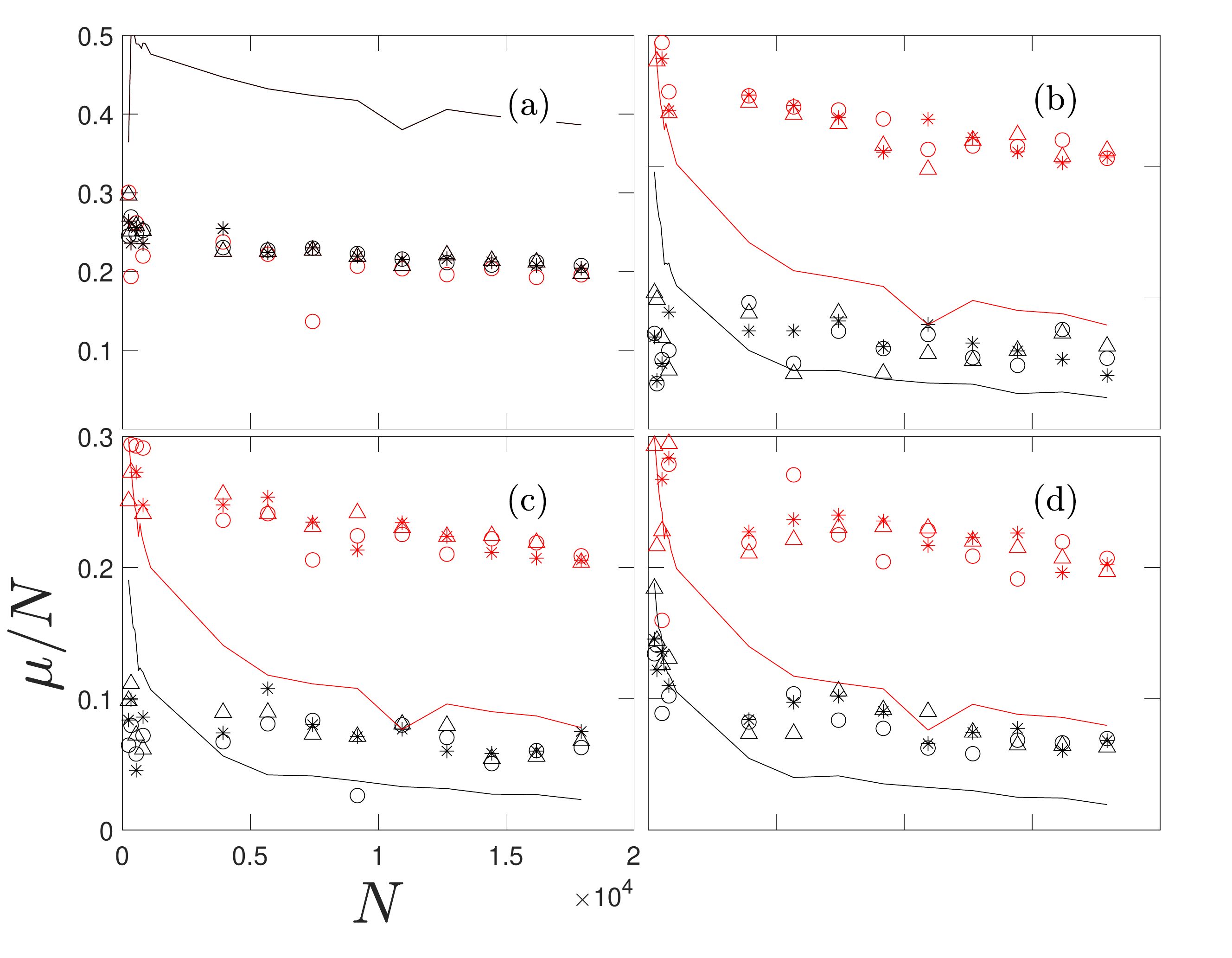}
\caption{(Color online) $\mu/N$ as a function of $N$ for the closed (different $y$ axis range 
than the other panels) (a), partially open with $R=0.1$ (b), $R=0.05$ (c), 
and completely open (d) tribaker map. 
The symbols $*$, $\triangle$ and $\circ$  stand for the 6th, 10th and 16th eigenstates, 
respectively.
The gray (red) points correspond to the measure  calculated for $\tilde{{\cal H}}^{R}_i$, 
while the black points for $\tilde{h}_i$. Full lines show the values 
of $\mu$ for the Husimi averages 
${<\!{\cal H}^{R}\!>}_j$ and the quantum repellers $Q_j$. 
}
\label{fig2}
\end{figure}

Next, we consider $\mu/N$ which is shown in Fig.~\ref{fig2} and is related to the fraction 
of the phase space corresponding to the given scaled resonance.
Again, as expected, there is essentially no difference here between the values for 
$\tilde{{\cal H}}^{R}_i$ and $\tilde{h}_i$ in the closed scenario. 
However, the localization is now better detected than with the $\Sigma$ 
measure for finite reflectivities and also for the completely open case. 
In fact, though the latter seems to be more localized, the results for partial openings 
are almost at the same level than this limiting case. 
Moreover, a greater localization with respect to the closed case is also detected in 
the Husimi distributions of the right resonances not only for the LR representation, 
which nevertheless is always much more localized. 
Finally, this localization seems to survive in the semiclassical limit (i.e., $\mu/N$ 
does not grow with $N$), 
but we leave a deeper investigation of this point for future work. 
Notice that the Husimi averages and the quantum repellers are more localized than individual 
scaled eigenfunctions, indicating the expected delocalization effect induced by this scaling 
on the original resonances.
%

\begin{figure}[b]
\includegraphics[width=0.90\columnwidth]{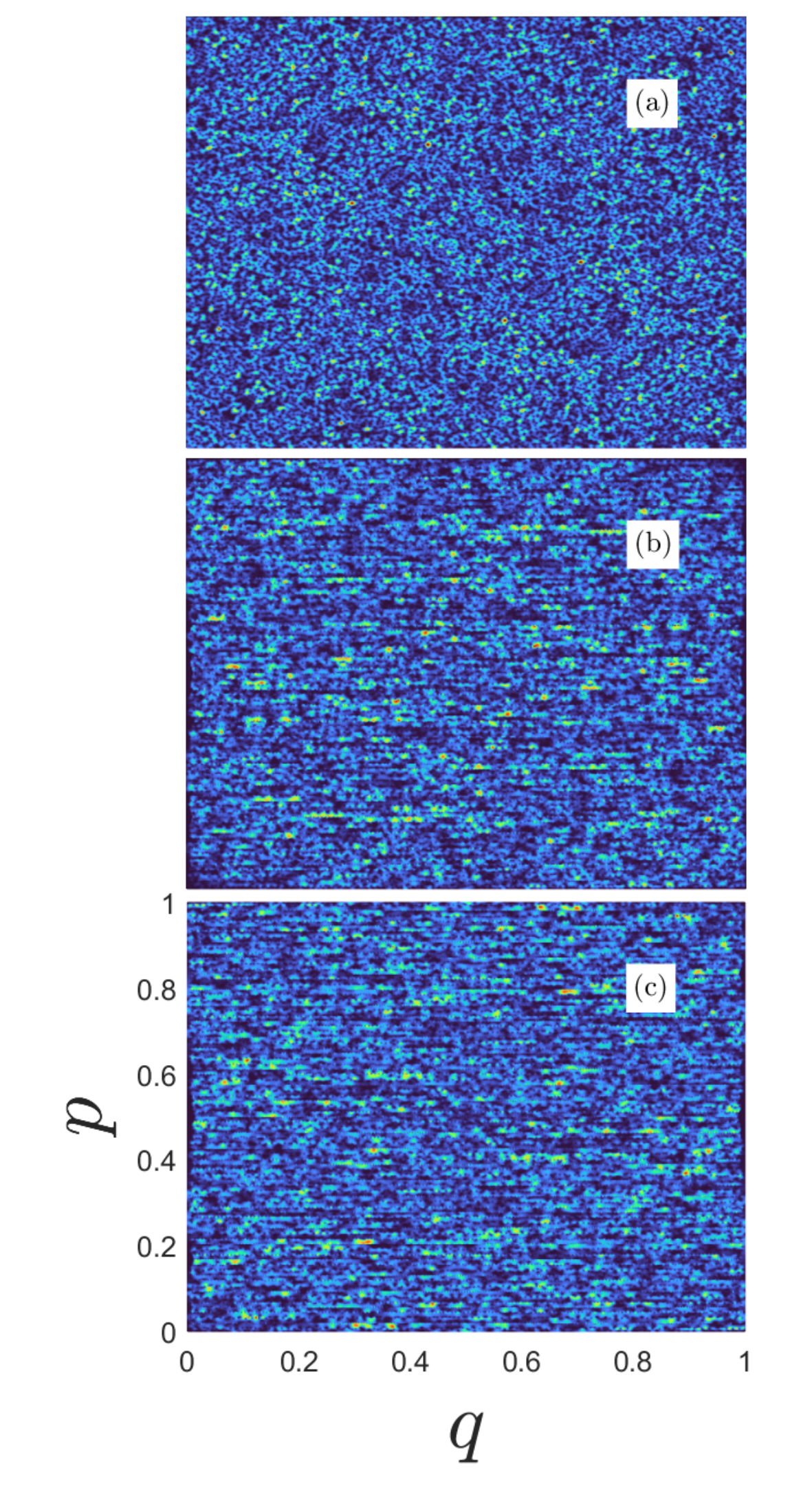}
\caption{(Color online) $\tilde{{\cal H}}^{R}_{10}(q,p)$ with $N=16185$ for the closed (a), 
partially open with $R=0.1$ (b), and $R=0.05$ (c). 
Lower to higher values go from blue (black) to red (gray).
}
\label{fig3}
\end{figure}

\begin{figure}
\includegraphics[width=0.90\columnwidth]{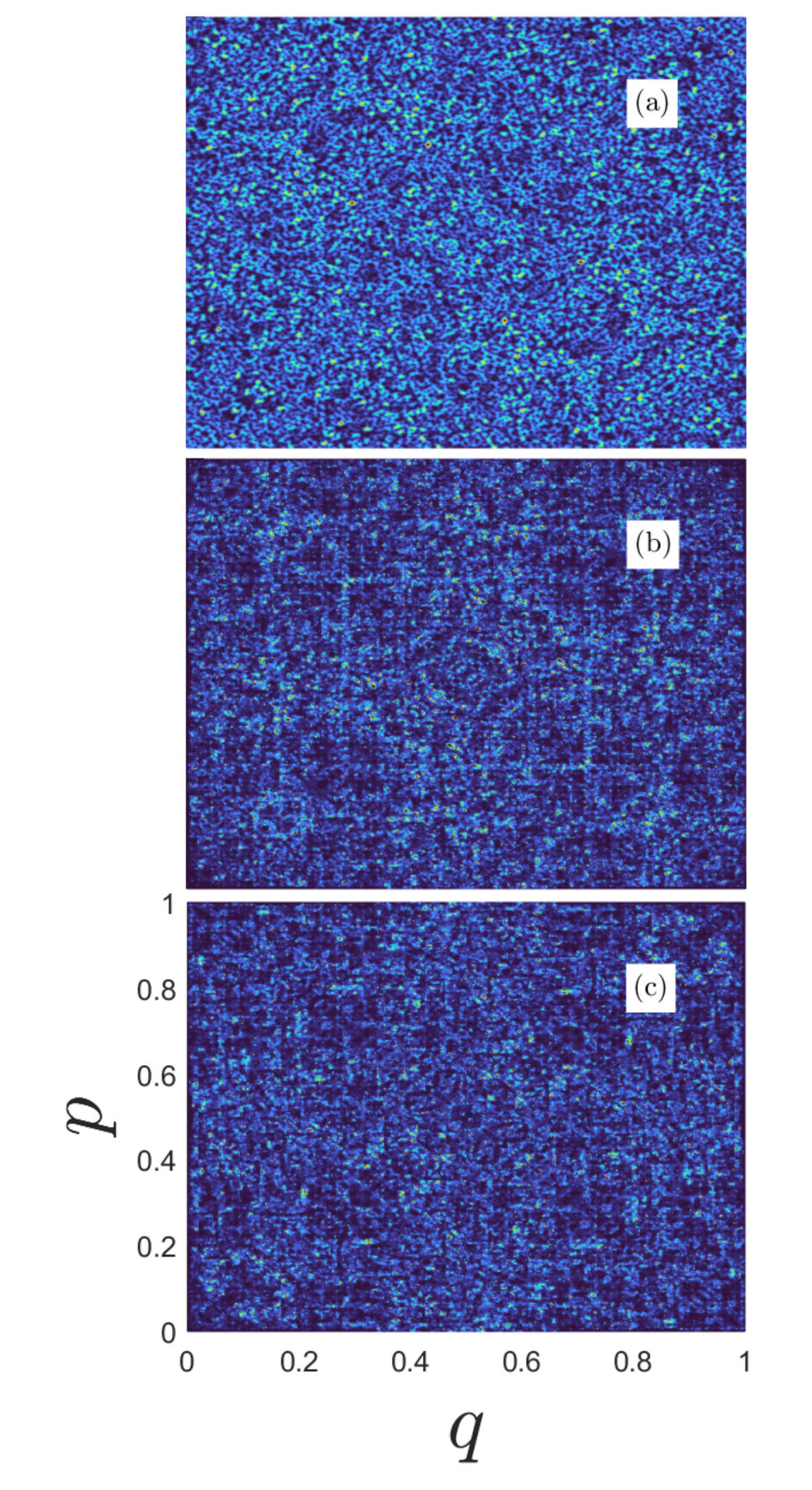}
\caption{(Color online) $\tilde{h}_{10}(q,p)$ with $N=16185$ for the closed (a), partially open with $R=0.1$ (b), 
and $R=0.05$ (c). Lower to higher values go from blue (black) to red (gray).
}
\label{fig4}
\end{figure}

We now turn to analyze some specific examples. 
In Fig.~\ref{fig3} we show the phase space representation $\tilde{{\cal H}}^{R}_{10}(q,p)$ 
for the largest $N$ evaluated in this work at all the values of $R$ considered 
in our calculation (a detailed explanation is given in the caption). 
No noticeable localization can be identified with the naked eye due to the large size 
of the Hilbert space. 
%
The same happens if we consider the LR representation $\tilde{h}_{10}(q,p)$ 
displayed in Fig.~\ref{fig4} for the same values of the parameter used in Fig.~\ref{fig3}.
We mention here that for the completely open case, some values of the scaled distributions 
(which were excluded from statistics) oscillate violently in the opening due to the 
extremely low values of the original resonances in that area. 
This also shows that scaling by the average in order to  analyze the properties of 
resonances does not have a smooth transition to the limiting completely open case.
%

We now provide clear evidence of localization on POs by means of directly looking at the 
phase space representations of eigenfunctions. 
We select a Hilbert space of a smaller size ($N=3936$) 
in order to make the identification of the features of phase space distributions easier. 
%
\begin{figure}
\includegraphics[width=1.1\columnwidth]{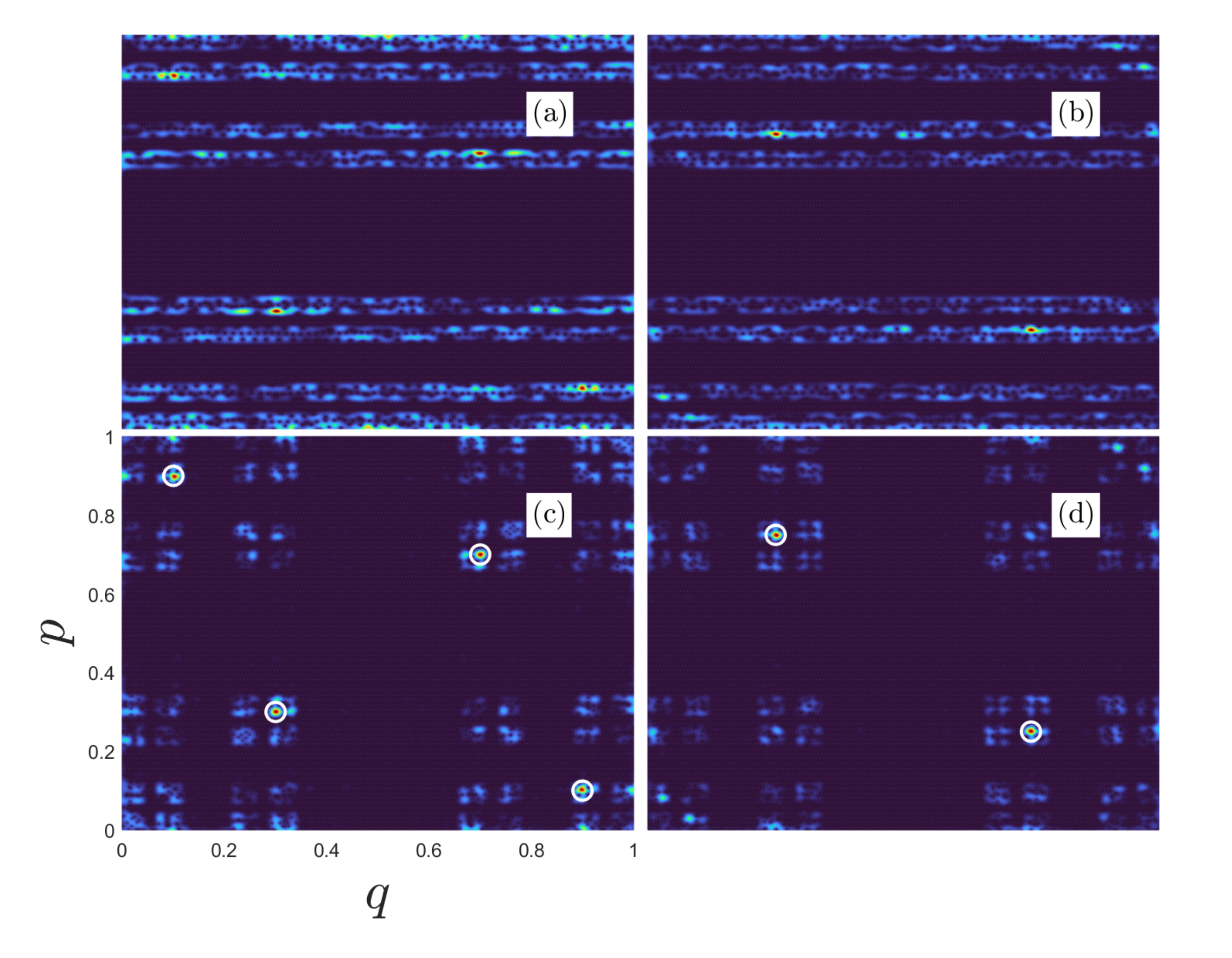}
\caption{(Color online) ${\cal H}^{R}_{18}(q,p)$ (a), ${\cal H}^{R}_{23}(q,p)$ (b), $h_{18}(q,p)$ (c), and 
$h_{23}(q,p)$ (d). 
Lower to higher values go from blue (black) to red (gray). 
Fixed points of short POs of period 4 and 2 are marked by means of white circles in (c) and (d).
In all cases we have taken $N=3936$ and $R=0.05$.
}
\label{fig5}
\end{figure} 
In Fig.~\ref{fig5} we show the non-scaled Husimi 
representation for the $18th$ and $23rd$ eigenfunctions (in decreasing eigenvalue modulus order as usual),
and the LR representations of it (see details of the different panels in the caption). 
In it, we can identify scarring by two short POs of period $4$ and $2$ respectively (whose 
fixed points are marked by means of white circles for the LR representation case). 
It is important to notice that we use a grid of $500$x$500$ points on the torus (at a reasonable 
computational cost) where 
coherent states are located in order to construct the distribution functions. 
The remarkable localization we have found not only proves that this happens on ordinary POs but that 
even at high $N$ some eigenfunctions can be dominated by a single scar function, as defined in the 
context of the short POs semiclassical theory \cite{art6,art7}. 
When comparing with the scaled version of the distributions in Figs. \ref{fig6} and \ref{fig7}, 
it becomes clear that the localization on POs is masked. 
%
\begin{figure}[b]
\includegraphics[width=0.90\columnwidth]{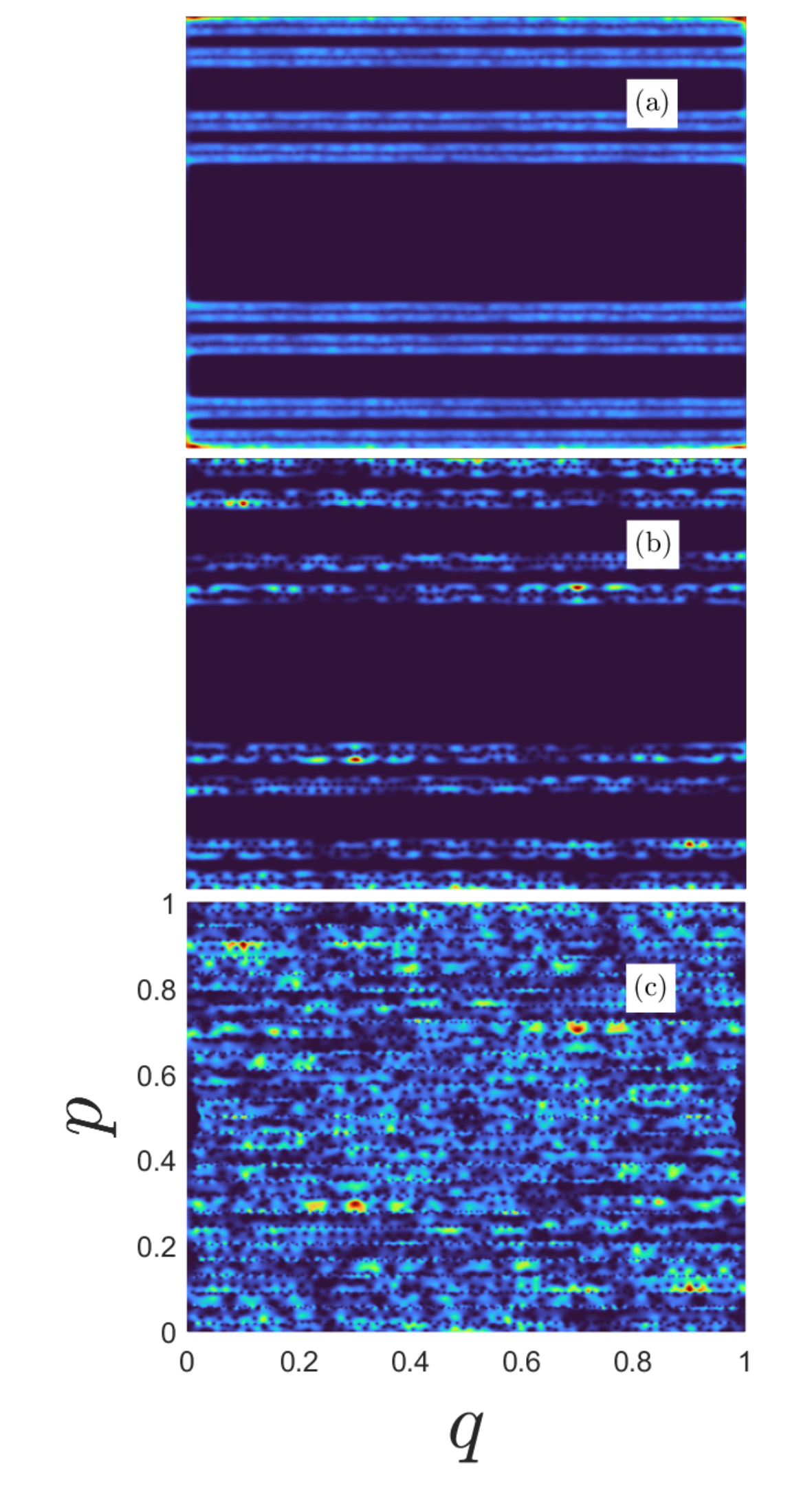}
\caption{Color online) ${<\!{\cal H}^{R}\!>}_j$ (a), ${\cal H}^{R}_{18}(q,p)$ (b), 
and $\tilde{{\cal H}}^{R}_{18}(q,p)$ (c). 
Lower to higher values go from blue (black) to red (gray). 
In all cases we have taken $N=3936$ and $R=0.05$.
}
\label{fig6}
\end{figure} 
The scaling dominates over the features of individual non-scaled 
eigenfunctions and the fluctuations that this procedure introduces translate into a localization of 
changing morphology, depending on the kind of representation used.
%
\begin{figure}
\includegraphics[width=0.90\columnwidth]{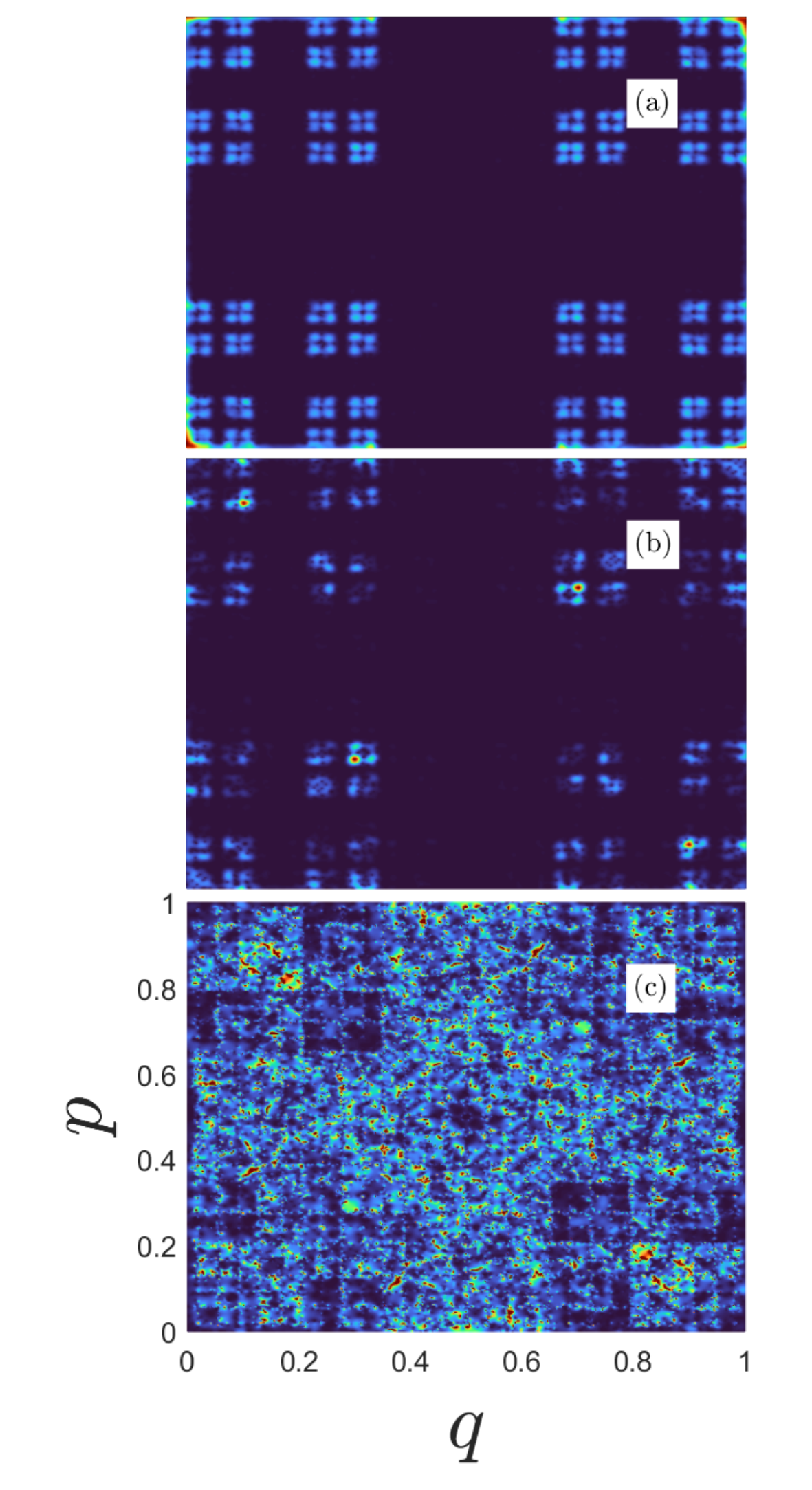}
\caption{(Color online) $Q_j$ (a), $h_{18}(q,p)$ (b), and $\tilde{h}_{18}(q,p)$ (c). 
Lower to higher values go from blue (black) to red (gray). 
In all cases we have taken $N=3936$ and $R=0.05$.
}
\label{fig7}
\end{figure} 

\section{Conclusions}
 \label{sec:Conclusions}

In this work we have studied the localization properties of scaled phase space distributions 
of resonances corresponding to the partially open tribaker map. 
We have found that for the longest lived ones, which define the quantum 
repeller, localization is present and persistent as $N$ grows. 
This is hinted by deviations measured by $\Sigma$ with respect to a conjectured universal intensity 
statistics \cite{Ketz3} when applied to the scaled LR representations. This representation is 
inspired by the spectral decomposition of the non unitary map and is the quantum analogue of 
the classical invariant set (moreover, it is the \textit{complete} analogy of the usual 
Husimi functions in closed systems, to which it converges as the partial 
opening becomes closer to $R=1$). This way of looking at resonances is an alternative to the 
configuration or the Husimi representations to which the conjectured universality does not apply. 
The existence of a general behavior for this representation is an interesting subject for future research, 
but we underline that as the quantum repeller is a coherent sum of long lived resonances its 
shape would be non-trivial.

Going into the details, we could find no appreciable differences with respect 
to a universal exponential distribution \cite{Ketz3} 
when evaluating $\Sigma$ for the scaled Husimis of the right resonances (as previously said, 
this is not the case for the LR representation). However, the norm ratio 
indicates more localization than in the closed system for the two phase space distributions, 
in particular for the LR representation. When looking at the phase space pictures we could see that localization of Husimi functions happens at different places than that of LR representations (both of scaled 
resonances), and that also in the latter case it lies outside of the region corresponding 
to the completely open repeller. In contrast, localization of the non-scaled resonances can 
be found on the same short POs (belonging to the repeller) in both, Husimis and LR representations. 
This leads us to conclude that although being a very useful tool, the scaling procedure must be modified 
in order to incorporate these findings. 

Consistently with the Shnirelman's theorem \cite{Shnirelman}, some recent studies suggest 
that the scarring persists in the semiclassical limit \cite{Vergini} for closed systems.  
To settle the very interesting question about the survival and amount of POs localization for the 
(partially) open case in the semiclassical limit, more investigations need to be carried out and we 
leave it for future work. But for that purpose we think it is fundamental to consider the quantum 
counterparts of both, the unstable and stable manifolds, being this a feature of the LR representation 
and the quantum repeller $Q_j$ by definition. In our view, it is in this sense that localization should 
be regarded in open systems, no matter if they are of partial or complete nature. A promising new 
representation could be of critical help in this effort \cite{HusimisNew}.

\section{Acknowledgments}
Support from CONICET is gratefully acknowledged.
This research has also been partially funded by the Spanish Ministry of Science, 
Innovation and Universities, Gobierno de Espa\~na under Contract No.~PID 2021-122711NB-C21.

%
\end{document}